\documentclass[aps,prd,twocolumn,superscriptaddress,nofootinbib,eqsecnumm]{revtex4}
\usepackage[pass,letterpaper]{geometry}
\usepackage{graphicx,color}
\usepackage{latexsym,amsmath,amsfonts,amssymb,booktabs}
\usepackage[pdfencoding=auto]{hyperref}
\usepackage{url}

\definecolor{MyBlue}{rgb}{0.15,0.15,0.70}
\hypersetup{
colorlinks=true,
citecolor=MyBlue,
linkcolor=MyBlue,
urlcolor=MyBlue
}

\newcommand{\gr}[1]{\boldsymbol{#1}}
\newcommand{\re}{{\rm e}}

\newcommand{\cH}{\mathcal{H}}

\newcommand{\be}{\begin{equation}}
\newcommand{\ee}{\end{equation}}
\newcommand{\beq}{\begin{equation}}
\newcommand{\eeq}{\end{equation}}
\newcommand{\bea}{\begin{eqnarray}}
\newcommand{\eea}{\end{eqnarray}}

\newcommand{\obs}{_{\rm O}}

\newcommand{\Gal}{_{\rm G}}

\newcommand{\dd}{\text{d}}

\usepackage[stable]{footmisc}

\newcommand\ees{\end{eqnarray}}
\newcommand\bees{\begin{eqnarray}}

\begin{document}

\title{A unified view of anisotropies in the astrophysical gravitational wave background}

\author{Cyril Pitrou}
\affiliation{Institut d'Astrophysique de Paris, CNRS UMR 7095,
 98 bis Boulevard Arago, 75014 Paris, France}

\author{Giulia Cusin}
\affiliation{Astrophysics Department, University of Oxford, DWB, Keble Road, Oxford OX1 3RH, UK}

\author{Jean-Philippe Uzan}
\affiliation{Institut d'Astrophysique de Paris, CNRS UMR 7095, 
98 bis Boulevard Arago, 75014 Paris, France}
\date{\today}

\begin{abstract}
In the literature different approaches have been proposed to compute
the anisotropies of the astrophysical gravitational wave
background. The different expressions derived, although starting from
our work \emph{Cusin, Pitrou, Uzan, Phys.Rev.D96, 103019 (2017)}
\cite{Cusin:2017fwz}, seem to differ. This article compares the
various theoretical expressions proposed so far and  provides a
separate derivation based on a Boltzmann approach. We show that all the theoretical formula in the literature are  
equivalent and boil down to the one of Ref.~\cite{Cusin:2017fwz} when a proper
matching of terms and integration by parts are performed. The  
difference between the various predictions presented for anisotropies
in a cosmological context can only lie in the astrophysical modeling of
sources, and neither in the theory nor in the cosmological description
of the large scale structures. Finally we comment on the gauge invariance of expressions. 

\end{abstract}
\maketitle

\section*{Introduction}

The anisotropic stochastic gravitational-wave (GW) background, generated by the superposition of various unresolved astrophysical and cosmological sources, has attracted a growing attention in the past years. It is probable  that the astrophysical gravitational wave background (AGWB) from unresolved stellar-mass binaries may be detected within a few years of operation of the LIGO-Virgo network~\cite{Abbott:2017xzg,2019arXiv190302886T}.

While the homogeneous component of the AGWB, i.e. its monopole, in a
perfectly homogeneous and isotropic spacetime has been studied for
many years, the computation of its anisotropies and their angular
power spectrum has only been derived recently. The computation of the
anisotropies of the AGWB relies on {\it (i)} the underlying cosmology
(assumed to be well described by a Friedmann-Lema\^{\i}tre cosmology)
{\it (ii)} the large scales structure or galaxy clustering and its effect on GW propagation (described using 
linear perturbation and effectively  including non-linearities in the matter evolution) and {\it
  (iii)} the local astrophysics on sub-galactic scales (given by the astrophysical modeling of the time-dependent GW luminosity
$\mathcal{L}_{\Gal}$ of a given galaxy as a function of halo mass
$M_{\Gal}$ and GW frequency at emission $\nu_{\Gal}$).

The first theoretical expression for the AGWB energy density
anisotropies was derived in Refs.~\cite{Cusin:2017fwz, Cusin:2017mjm}, using a
coarse-graining approach,  \textcolor{black}{that is distinguishing
  different physical scales in the problem. The largest scale is the cosmic
  flow scale, which controls the matter overdensity and
  velocity. Galaxies are then assumed to be a biased tracer of the total matter (including dark matter),
  and we associate to each galaxy an effective GW luminosity (in the
  galaxy rest frame). It takes into account the contribution of
  different astrophysical GW sources as a function of galaxy physical parameters such as masses and metallicities.} In Refs.~\cite{Cusin:2017fwz, Cusin:2017mjm} both a 
covariant expression valid in any spacetime and its application to the case of a perturbed Friedmann-Lema\^{\i}tre
universe are presented. Using this framework,  the first predictions of the
AGWB power spectrum have been presented in Ref.~\cite{Cusin:2018rsq}, for the contribution of binary black
holes (BH) mergers and for frequencies in the LIGO-Virgo band. Making use of  the astrophysical framework
described in Refs.~\cite{Dvorkin:2016okx,Dvorkin:2016wac,2018MNRAS.479..121D}, the
influence of various astrophysical parameters/functions (BH formation
models, BH mass cut-off, stellar initial mass function, metallicity)
on the angular power spectrum was investigated in
Ref.~\cite{Cusin:2019jpv} and extended to the LISA band in
Ref.~\cite{Cusin:2019jhg}. Similar predictions were proposed, on the
basis of the same formalism, in Refs.~\cite{Jenkins:2018uac,Jenkins:2018kxc} (see Ref.~\cite{Cusin:2018ump} for a comment on the analytic approach used in these works).  
The first attempt to describe anisotropies of a GW with a Boltzmann approach was proposed by Ref.~\cite{Contaldi:2016koz} while Ref.\,\cite{Cusin:2018avf} refines it by introducing an emissivity function that realistically describes GW emission at the galactic scale. A systematic study of cross-correlation with the galaxy distribution is presented in Ref.\,\cite{Alonso:2020mva}. 

Recently, Ref.~\cite{Bertacca:2019fnt} proposed a new derivation  of
  anisotropies in a cosmological context, starting from the covariant
  expression presented in Ref.~\cite{Cusin:2017fwz}. Using methods
  adapted to large scale structure observed in redshift space, the
  authors claim the presence of additional \textcolor{black}{projection} effects in their result, not
  previously taken into account in the literature. \textcolor{black}{Projection effects
  arise in general when indirectly reconstructing distances of astrophysical
  objects from redshifts. In particular they argue that a ``Kaiser
  projection term'', due to peculiar velocity Doppler shifts, would
  become relevant for large angular separations. However, as we shall
  show further, this effect is only present when having access to the
  source location, which is not the case for a background
  like the AGWB. Indeed, a background depends only on the
  direction of observation, and this is reflected in the theoretical expression of its anisotropy which
  is obtained integrating along the line of sight.}

The goal of this article is to compare different expressions for anisotropies,  obtained
using the covariant approach of Ref.~\cite{Cusin:2017fwz} as a starting point and making use of different perturbation methods.  In particular we investigate the origin
of the new projection terms recently found in Ref.~\cite{Bertacca:2019fnt}.  
To that purpose, we will first provide in \S~\ref{sec2} a line-of-sight approach of our
covariant expression~\cite{Cusin:2017fwz}. Then, after a brief reminder of its implementation within cosmological perturbation
theory in \S~\ref{sec3}, we compare in \S~\ref{sec4} the results of
Ref.~\cite{Cusin:2017fwz} in a cosmological context with the result
recently presented in Ref.~\cite{Bertacca:2019fnt}.  We show that the
discrepancy arises from the different choices on how to model the GW
luminosity of galaxies. In particular we show that projection effects,
which play an important role in galaxy surveys, do not correspond to observable 
cosmological effects in the context of a background. 
To finish, we comment on the modeling of the luminosity perturbation in \S~\ref{SecL}.\\

\noindent{\bf Notation:} We assume a standard Friedmann-Lema\^{\i}tre background spacetime with scale factor $a$, conformal time $\eta$ and conformal Hubble function $\cH\equiv \dd\ln a/\dd\eta$.

\section{Line of sight expression}\label{sec2}

A simplified approach is to describe the AGWB observed today in terms
of its energy density per units of observed frequency $\nu_{\obs}$ and
solid angle $\dd \Omega_{\obs}$, as 
\be\label{Master1}
 \frac{\dd^3\rho_{\rm GW}(\gr{e}, \nu_{\obs})}{\dd\nu_{\obs}\dd^2\Omega_{\obs}}=\int
 \frac{n_{\Gal} [x^\mu(\lambda)]}{\left[1+z(\lambda)
   \right]^3}\,\frac{\mathcal{L}_{\Gal}[x^\mu(\lambda),\nu(\lambda)]}{4\pi} \frac{\dd
   \tau}{\dd \lambda}\dd\lambda \,,
 \ee
where $x^\mu(\lambda)$ is the  geodesic followed by the GW observed in direction $\gr{e}$ and $z(\lambda)$  describes the redshift along that geodesic. It is defined as $1+z(\lambda) \equiv  [u_\mu p^\mu_{\rm GW}]_{x^\mu(\lambda)}/[u_\mu p^\mu_{\rm GW}]_{\obs}$, where $p^\mu_{\rm GW}=\dd x^\mu/\dd \lambda$ is the GW four-momentum and $u^\mu_{\obs,\Gal}$ the tangent vector to the observer ($\text{\small O}$) or source ($\text{\small G}$) geodesic. The physical number density of galaxies is denoted as  $n_{\Gal}$ and the GW luminosity of galaxies as ${\cal L}_{\Gal}$. They are both defined in the matter comoving frame (i.e. comoving with $u^\mu_{\Gal}$ at $x^\mu(\lambda)$). The GW luminosity depends on the redshifted frequency $\nu(\lambda) = \nu_{\obs} (1+z(\lambda))$.

The expression~(\ref{Master1}) is fully covariant in the sense that it
is valid for any spacetime metric. It was derived using the distance
reciprocity relation $D_{\rm A}/D_{\rm L}=1/(1+z)^2$ between the
luminosity ($D_{\rm L}$) and angular ($D_{\rm A}$) distances, since
the power received scales by definition as $1/D_{\rm L}^2$ while the
area spanned by a solid angle scales like $D_{\rm A}^2$. The factor
$\dd\tau/\dd \lambda = -u_\mu p^\mu_{\rm GW}$ accounts for the
physical depth spanned (equal to the proper time of the source spanned
$\dd \tau$) in an infinitesimal change $\dd \lambda$ of the affine
parameter used in the geodesic integration. Instead of integrating with this affine parameter, one usually integrates over the conformal time $\eta$ and makes use of the relation $(\dd \tau/\dd \lambda) \dd \lambda = (\dd \tau/\dd \eta)\dd \eta$.

Equation~\eqref{Master1} is rather formal and requires some extra
manipulations to obtain the general expression in a  perturbed cosmological framework. However, in this general form it is apparent that it is
the integral form of a Boltzmann equation with a source term  but
without any collision term (as was actually initially proposed in the
context of GW by Ref.~\cite{Contaldi:2016koz}). Indeed, defining a
total derivative in phase space along a geodesic by $\frac{\rm D}{{\rm D} \eta} \equiv \partial_\eta + \frac{p^i}{p^0}\partial_i +
\frac{\dd p^i}{\dd \eta}\frac{\partial}{\partial p^i}$, Eq.~\eqref{Master1} is the integrated version of
\be\label{Boltzmann}
\frac{\rm D}{{\rm D} \eta} f(x^\mu,\nu,\gr{e}) = \mathcal{E}(x^\mu,\nu,\gr{e})\,,
\ee
which is a Boltzmann-type equation for the distribution function
\be
f(x^\mu,\nu,\gr{e})  \equiv \nu^{-3} \frac{\dd^3 \rho_{\rm GW}(x^\mu,\nu,\gr{e})}{\dd \nu\dd^2 \Omega}\,,
\ee
with emissivity function 
\be\label{emissivity}
\mathcal{E}(x^\mu,\nu,\gr{e})\equiv  \frac{\dd
  \tau}{\dd \eta}(x^\mu,\gr{e}) n_{\Gal}(x^\mu) \frac{{\cal L}_{\Gal}(x^\mu,\nu)}{4 \pi \nu^3}\,.
 \ee
The galaxy number density $n_{\Gal}$ converts the galaxy luminosity ${\cal L}_{\Gal}$ into a luminosity per unit of volume. In this Boltzmann approach, the factor $\dd \tau/\dd \eta$ accounts for the fact that the rate of emission is defined, as it should, with respect to the time in the comoving frame of GW sources. The practical expression of this time conversion factor is
\be\label{TimeConversion}
\frac{\dd \tau}{\dd \eta} = \frac{-u_\mu p^\mu}{p^0} = (1+z) \frac{\nu_{\obs}}{p^0}\,.
\ee
Neither the line of sight expression \eqref{Master1} nor the Boltzmann
equation \eqref{Boltzmann}  make any hypothesis on the spacetime
symmetries. Note however that at the core of these descriptions is
indeed hidden some kind of a coarse-graining since the source term
involves the continuous galaxy number density field, which allows one
to define a continuum of sources. All the works mentioned earlier make 
such an assumption, whether it is explicitly said or not. For
comparison, we stress that this is not the case for the 
collision term of the CMB radiative transfer, which is based on the more fundamental microscopic form of Compton interactions.

\section{Linear perturbation theory}\label{sec3}

\subsection{General expression}

Let us now restrict to the framework of linear cosmological perturbation theory, with a metric perturbed in full generality (but
with only scalar perturbations describing the large scale structure) as
\begin{align}\label{FL}
\dd s^2&=a^2\left[-(1+2\psi)\dd\eta^2+2\partial_ i B \dd x^i \dd \eta\right.\nonumber\\
  &\left. \quad+(1-2\phi)\delta_{ij}\dd x^i\dd  x^j+2\partial_i \partial_j E \dd x^i \dd x^j    \right]\,.
\end{align}
One needs to solve for the geodesic of gravitons, so as to determine their trajectory,  and  the frequency evolution and the redshift along a geodesic. In practice, when integrating with the parameter $\eta$, it is sufficient to consider the background geodesic which is a straight line, since time-delays and lensing appear only as second order contributions. For simplicity we omit here the contributions of the perturbations at the observer's position since observables do not depend on them. The redshift perturbation is
\bea
1+z &=& (1+\bar z) [1 + \delta \ln (1+z)]\,,\\
\delta \ln (1+z)&\equiv&-\psi + e^i \partial_i (v+B) \label{deltalnz} \\
&-&\int_\eta^{\eta_{\obs}} [\psi'+\phi' +e^i e^j\partial_i \partial_j(B-E')]\dd \tilde \eta\,,\nonumber
\eea
where $1+\bar z \equiv 1/a$, and $u^i=a^{-1}v^i=a^{-1}\partial^i v$ is the spatial component of the matter velocity. The perturbed time conversion factor [Eq.~\eqref{TimeConversion}] is
\be\label{TimeConversion2}
\frac{\dd \tau}{\dd \eta} = a[1+\psi+e^i \partial_i (v+B)]\,.
\ee
Hence, using $\bar \nu = (1+\bar z )\nu_{\obs}$, we find from Eq.~\eqref{Master1} that the linearly perturbed GW background energy density per units  of observed frequency and angle is 
\begin{widetext}
\begin{align}\label{MasterGeneral}
\frac{\dd^3\rho_{\rm     GW}(\gr{e}, \nu_{\obs})}{\dd\nu_{\obs}\dd^2\Omega_{\obs}}&=\int^{\eta_{\obs}} \dd  \eta a^4  \bar n_{\Gal} \frac{\bar{\mathcal{L}}_{\Gal}(\eta,\bar
   \nu)}{4\pi} \times  \Big[1+\delta_{\Gal}+\delta_{\cal L}+\left(\frac{\partial \ln    \bar{\mathcal{L}}_{\Gal}}{\partial \ln \bar \nu}-2\right) e^i \partial_i (v+B) 
   +\left(4-\frac{\partial \ln
    \bar{\mathcal{L}}_{\Gal}}{\partial \ln \bar \nu}\right) \psi
  \nonumber\\
  &\qquad \qquad\qquad \qquad\qquad \qquad +\left(3-\frac{\partial \ln
    \bar{\mathcal{L}}_{\Gal}}{\partial \ln \bar \nu}\right)    \int_\eta^{\eta_{\obs}} [\psi'+\phi' + e^i e^j\partial_i \partial_j(B-E')] \dd \tilde \eta \Big]\,,
\end{align}
\end{widetext}
where the luminosity perturbation is defined by ${\cal L}_{\Gal}(\eta,\nu) =
\bar{{\cal L}}_{\Gal}(\eta,\nu) [1+ \delta_{\cal L}(\eta,\nu)]$, and  the integration
is performed on the background geodesic $x^i = e^i
(\eta_{\obs}-\eta)$. The terms with $\partial \ln
\bar{\mathcal{L}}_{\Gal}/\partial \ln \bar \nu$  come by Taylor expanding ${\cal
  L}_{\Gal}(\nu, \eta)$ around $\bar\nu$. What remains to be
specified, is the relation between the galaxy density contrast
$\delta_{\Gal}$ and the underlying cosmological perturbations, and
also  a model for the luminosity perturbations $\delta_{\cal L}$.

\subsection{Gauge invariance}

Under a general gauge transformation generated by the vector field $\xi^\mu=(T,\partial^i L)$, the perturbations transform as $\psi
\overset{{\rm GT}}{\to}
\psi + \cH T + T'$, $\phi \overset{{\rm GT}}{\to} \phi - \cH T$, $E \overset{{\rm GT}}{\to} E +L$, $B \overset{{\rm GT}}{\to} B +L'-T$, $v \overset{{\rm GT}}{\to} v-L'$,  $\delta_G \overset{{\rm GT}}{\to} \delta_G + (\ln \bar n_G)'
T$ and  $ \delta_{\cal L} \overset{{\rm GT}}{\to} \delta_{\cal L} + T \partial
\ln \bar{\cal L}_{\Gal}/\partial \eta$. It follows that two of the
four scalar perturbations ($\phi,\psi,B,E$) can always be set to zero by a proper choice of $(T,L)$. Note that ${\cal L}_G$ is defined as the luminosity  seen in the matter comoving frame, that is with respect to a tetrad  field whose timelike vector is the matter velocity. This differs from  the standard approach of the Boltzmann equation, where the distribution function, and thus its sources and associated collision term, are defined with respect to a tetrad whose timelike vector is proportional to $(\dd
  \eta)_\mu$~\cite{Pitrou:2008hy}. The perturbations on these two different time slices are related by
  $\delta_{\cal L}^{\{\dd \eta\}}  = \delta_{\cal L} +e^i \partial_i(v+B)\partial\ln \bar{\cal
    L}_{\Gal}/\partial \bar \nu$, and $\delta_{\cal L}^{\{\dd \eta\}}
  $ transforms under a general gauge transformation similarly to $\delta_{\cal
    L}$, but with an additional term $-e^i\partial_i T \partial\ln
  \bar{\cal L}_{\Gal}/\partial \bar \nu$
  \cite{Durrer:1993db,Durrer:2001gq,Pitrou:2007jy}. Using these transformation properties, it is easy to check 
  that the general expression~\eqref{MasterGeneral} is gauge invariant, as it should be the case since it describes an observable quantity.

\subsection{Gauge fixing}

Restricting to the Newtonian gauge (NG) by setting $B=E=0$ in Eq.\,(\ref{MasterGeneral}),  we get
\begin{align}\label{Master2}
 &\left.\frac{\dd^3\rho_{\rm     GW}(\gr{e}, \nu_{\obs})}{\dd\nu_{\obs}\dd^2\Omega_{\obs}}\right|_{\rm NG}=\int^{\eta_{\obs}} \dd  \eta a^4  \bar n_{\Gal} \frac{\bar{\mathcal{L}}_{\Gal}(\eta,\bar
   \nu)}{4\pi} \times\nonumber\\
  &\Big[1+\delta_{\Gal}+\delta_{\cal L}+\left(4 -\frac{\partial \ln
    \bar{\mathcal{L}}_{\Gal}}{\partial \ln \bar \nu} \right)\psi+\left(\frac{\partial \ln
    \bar{\mathcal{L}}_{\Gal}}{\partial \ln \bar \nu}  -2\right) e^i
    \partial_i v \nonumber\\
  &+\left(3 -\frac{\partial \ln
    \bar{\mathcal{L}}_{\Gal}}{\partial \ln \bar \nu}\right)   \int_\eta^{\eta_{\obs}} (\psi'+\phi' )\dd \tilde \eta \Big]\,.
\end{align}
In Eq.~(4) of Ref.~\cite{Cusin:2018rsq}, which was derived in Newtonian gauge, we introduced a phenomenological bias factor to relate $\delta_{\Gal}$ to
matter overdensities (in practice relating the comoving density contrasts), and we set $\delta_{\cal L}=0$.

Even though the expression \eqref{Master2} is given in a specific (Newtonian) gauge, the perturbation variables can always be promoted
to gauge invariant variables replacing $\psi\to \psi +
\cH(B-E')+B'-E''$, $\phi\to \phi - \cH(B-E')$, $v \to v + E'$,
$\delta_{\Gal} \to \delta_{\Gal} + (B-E')(\ln \bar n_G)'$ and
$\delta_{\cal L} \to \delta_{\cal L} + (B-E')\partial \ln \bar{\cal
  L}_{\Gal}/\partial \eta$.  This standard method allows one  to re-express the result in an arbitrary gauge, and it is straightforward to check that after integration by parts of the integrated effects, one recovers the expression  \eqref{MasterGeneral} in an arbitrary gauge.

Similarly, restricting to the synchronous gauge (SC) (with comoving condition on cold matter) by setting $\psi=v=B=0$ in Eq.\,(\ref{MasterGeneral}), we get
\begin{align}\label{Master3}
& \left.\frac{\dd^3\rho_{\rm GW}(\gr{e}, \nu_{\obs})}{\dd\nu_{\obs}\dd^2\Omega_{\obs}}\right|_{\rm SC}=\int^{\eta_{\obs}} \dd
   \eta a^4  \bar n_{\Gal} \frac{\bar{\mathcal{L}}_{\Gal}(\eta,\bar
     \nu)}{4\pi}\times\\
&  \left[1+\delta_{\Gal}+\delta_{\cal L}+\left(\frac{\partial \ln
       \bar{\mathcal{L}}_{\Gal}}{\partial \ln \bar \nu}
     -3\right)\int_\eta^{\eta_{\obs}} (e^i e^j \partial_i \partial_j
                           E'-\phi')\dd \tilde \eta\right]\,,\nonumber
\end{align}
where the effect of metric perturbations appears as an integrated term. 

\section{Comparison with literature}\label{sec4}

Recently Bertacca {\em et.~al.}~\cite{Bertacca:2019fnt} (see their
Section~III, and we follow here the notation of v1 of this preprint) also focus on a perturbed Friedmann-Lema\^{\i}tre
spacetime and using the \emph{cosmic rulers
  formalism}~\cite{Schmidt:2012ne} find a general expression for the AGWB anisotropies on a perturbed cosmology. We start from
 Eq.~(81) of ~\cite{Bertacca:2019fnt} (in an arbitrary gauge) and for the sake of comparison we restrict it to the
Newtonian gauge. This gives [setting to zero perturbations at the observer's position] 
\begin{align}
\label{Bertacca81}
 &\left.\frac{\dd^3\rho_{\rm
   GW}(\gr{e}, \nu_{\obs})}{\dd\nu_{\obs}\dd^2\Omega_{\obs}}\right|_{\rm Bertacca} = \int^{\eta_{\obs}}
   \dd \eta a^4  \bar n_{\Gal}  {\mathcal K}(\bar z,  \nu_{\obs}(1+\bar z)) \nonumber\\
  &\Bigg\{ 1+ \delta_{\Gal} + \left(- b_\re  + 3 +  {\cH'\over \cH^2} -\frac{1}{\cH}\frac{\dd}{\dd \eta}\right) \psi \\
 &+ \left(b_\re -1 - {\cH'\over \cH^2}
   +\frac{1}{\cH}\frac{\dd}{\dd \eta}\right)  e^i \partial_i v \nonumber\\
  &+\frac{1}{\cH}(\psi'+\phi') -    \left(b_\re - 2 - {\cH'\over \cH^2}\right)
    \int_\eta^{\eta_{\obs}} (\phi'+\psi') \dd \tilde \eta \Bigg\}\,, \nonumber
 \end{align}
where the evolution bias is defined as  $ b_\re = \cH^{-1}\dd\ln(\bar
n_{\Gal} a^3)/\dd \eta$ and the function $ {\mathcal K}$ corresponds to $\bar{\cal L}_{\Gal}/(4 \pi)$ in our notation. This expression looks very different from our Eq.~\eqref{Master2} but it can be easily compared with it.
Using integration by parts,  and  noticing for instance that 
\be
a^4 n_{\Gal}\left(b_\re -\cH'/\cH^2+\frac{1}{\cH}\frac{\dd}{\dd \eta}\right) \psi =  a\frac{\dd}{\dd\eta} \left(\frac{a^3 n_{\Gal }}{\cH} \psi\right)\,,\nonumber
\ee
and replacing ${\cal K} \equiv \bar{\cal L}_{\Gal}/(4 \pi)$, Eq.\,(\ref{Bertacca81})
reduces to\footnote{Note that instead of using integration by parts to
  obtain Eq.~\eqref{MasterBertacca} from \eqref{Bertacca81}, we could consider that the time
variable in the integration of Eq.~\eqref{Bertacca81} is in fact
$\eta_z$, and then use the change of variable~\eqref{etaz} to
integrate on $\eta$. From first order expansion of all background
quantities in $\eta_z-\eta$, one recovers Eq.~\eqref{Master2} up to
the difference \eqref{Diff}.}
\begin{align}\label{MasterBertacca}
 &\left.\frac{\dd^3\rho_{\rm
   GW}(\gr{e}, \nu_{\obs})}{\dd\nu_{\obs}\dd^2\Omega_{\obs}}\right|_{\rm Bertacca}=\int^{\eta_{\obs}} \dd
   \eta a^4  \bar n_{\Gal} \frac{\bar{\mathcal{L}}_{\Gal}(\eta,\bar
     \nu)}{4\pi} \times \\
     &\Big[1+\delta_{\Gal}+\left(4-\frac{\partial \ln    \bar{\mathcal{L}}_{\Gal}}{\partial \ln \bar \nu} \right) \psi +\left(\frac{\partial \ln    \bar{\mathcal{L}}_{\Gal}}{\partial \ln \bar \nu} -2\right) e^i \partial_i v \nonumber\\
  &+\left(3-\frac{\partial \ln    \bar{\mathcal{L}}_{\Gal}}{\partial
    \ln \bar \nu} \right)   \int_\eta^{\eta_{\obs}} (\psi'+\phi') \dd
    \tilde \eta-\frac{\partial \ln
    \bar{\mathcal{L}}_{\Gal}}{\partial \eta}\frac{\delta \ln (1+z)}{\cH}\Big].\nonumber
\end{align}

We remark that the difference with Eq.~\eqref{Master2}  is the last term and the absence
of $\delta_{\cal L}$. Hence Eq.~\eqref{MasterBertacca} is equivalent
to Eq.~\eqref{Master2} with the choice $\delta_{\cal L}=(\eta_z-\eta) \partial\ln
\bar{\cal L}_{\Gal}/\partial \eta$. In order to gain insight on the
physical significance of this choice, let us define the luminosity perturbation on constant redshift hypersurfaces
\be\label{deltaZ}
\delta_{\cal L}^z  \equiv \delta_{\cal L} - (\eta_z-\eta) \frac{\partial\ln
\bar{\cal L}_{\Gal}}{\partial \eta}\,, 
\ee
where 
\be\label{etaz}
\eta_z \equiv \eta - \cH^{-1}\delta \ln (1+z)\,,
\ee
is the time at which a galaxy in a fiducial unperturbed cosmology would have the same redshift as a galaxy otherwise located
at $x^\mu(\eta)$ in the (true) perturbed cosmology. It follows that $\delta_{\cal L}=(\eta_z-\eta) \partial\ln
\bar{\cal L}_{\Gal}/\partial \eta$ corresponds to setting the perturbation of the galaxy luminosity to zero on constant redshift hyper surfaces $\delta_{\cal L}^z=0$, which is the choice of ~\cite{Bertacca:2019fnt} to describe the perturbation of the galaxy luminosity. The difference
between Eqs.~\eqref{Master2} and \eqref{MasterBertacca} is then reduced to
\begin{align}\label{Diff}
&\left.\frac{\dd^3\rho_{\rm
   GW}(\gr{e}, \nu_{\obs} )}{\dd\nu_{\obs}\dd^2\Omega_{\obs}}\right|_{\rm Bertacca}\\
               \
  &= \left.\frac{\dd^3\rho_{\rm GW}(\gr{e},\nu_{\obs})}{\dd\nu_{\obs}\dd^2\Omega_{\obs}}\right|_{\rm Cusin} - \int^{\eta_{\obs}} \dd   \eta a^4  \bar n_{\Gal} \frac{\bar{\mathcal{L}}_{\Gal}(\eta,\bar
     \nu)}{4\pi} \delta_{\cal L}^z\, \nonumber\,.
\end{align}

We showed how to perform this comparison in the Newtonian gauge for
simplicity. However,  using the same types of integrations by parts,
it is easy to check that the difference is the same in an arbitrary
gauge, i.e. the difference between Eq.~\eqref{MasterGeneral} and  Eq. (81) of Ref.~\cite{Bertacca:2019fnt} is also given by
Eq.~\eqref{Diff}. We stress that both these expressions are gauge invariant even though they differ, since the difference involves $\delta_{\cal L}^z$  which is itself a gauge invariant quantity, as $\delta \ln (1+z) \overset{{\rm
    GT}}{\to} \delta \ln (1+z) -\cH T $. It also follows that the
difference between the restriction to the synchronous gauge~\eqref{Master3} and Eq.~(85) of Ref.~\cite{Bertacca:2019fnt} is given by \eqref{Diff}. 

As a result, we find that the new terms found  in  Ref.~\cite{Bertacca:2019fnt} written in the form of typical projection terms in large scale structure observables [see for example the Kaiser-like term in the third line of Eq.~\eqref{Bertacca81}], can be removed with some integration by parts, hence do not correspond to measurable physical effects. This is actually a general result and applies to any background: projection effects, very relevant when dealing with large scale structure observables (i.e. sources that we can observe directly and localize in space) are not effective when dealing with a background. A background by definition is given by the superposition of signals from unresolvable sources and it is therefore \emph{blind} to the location in redshift space of sources, which need to be described using a local model for galactic physics.

\section{Luminosity perturbation}\label{SecL}

In Ref.~\cite{Bertacca:2019fnt}, the perturbation of the effective
luminosity as a function of the observed redshift is set to zero on
constant redshift hypersurfaces, i.e. $\delta_{\cal L}^z = 0$.  Since we do not directly observe the luminosity of galaxies but only the integrated flux that
we receive from them, in our work Ref.~\cite{Cusin:2017fwz} we had assumed the luminosity of a galaxy to be a function of its local time $\eta$ (even though it is only a coordinate and ideally one should prefer to use
its proper time), rather than a function of redshift which depends on
the way galaxies are observed later, and we had 
chosen an astrophysical model  of GW sources such that the luminosity
perturbation vanishes in the Newtonian gauge, i.e. $\delta_{\cal
  L}=0$. A similar choice for source luminosity perturbations is made
in the context of the cosmic infrared background~\cite{Tucci:2016hng}. Furthermore, constant redshift hypersurfaces are observer dependent, hence parametrizing the galaxy luminosity in terms of the observed
redshift \textcolor{black}{(as done in Ref.~\cite{Bertacca:2019fnt} for the AGWB)} may be useful and well-motivated in the case we could directly access and measure the emitted luminosity, which is not the case in the context of a background. 

\section*{Conclusion}

We have compared different predictions for the anisotropies of the
gravitational wave background proposed in the literature, starting
from our covariant approach of Ref.~\cite{Cusin:2017fwz}. We have
shown that they are all equivalent and they all contain the same type
of cosmological effects. Differences in predictions can only arise in the way galactic physics is computed.
In particular, the expression for the background anisotropies proposed in Ref.~\cite{Bertacca:2019fnt} reduces, after some integration by parts 
to the former expressions of Refs.~\cite{Cusin:2017fwz,Cusin:2017mjm},
except for a term [see Eq.~(\ref{Diff})] which is the result of a different modeling of GW luminosity as a function of redshift and frequency.  Moreover, we have shown that both the approaches of  Ref.~\cite{Bertacca:2019fnt} and Refs.~\cite{Cusin:2017fwz,Cusin:2017mjm} used to expand the covariant expression of Ref.~\cite{Cusin:2017fwz} on a perturbed cosmological framework,  give a gauge-invariant observable. This is a nice cross-check of the results derived so far: by construction, gauge invariance is built in the computation
of an observable, and different choices of the time coordinate used to parametrize it should not change its transformation property under gauge transformations.

More generally, any difference between AGWB power spectra can arise from {\it (i)} the
underlying cosmology ($\bar n_{\Gal}$) {\it (ii)} the large scales
structure or galaxy clustering ($\delta_{\Gal}$ and the bias model)
and/or {\it (iii)} the local astrophysics on sub-galactic scales
encompassed by the luminosity function ${\cal L}_{\Gal}(\eta,\nu)$ (its background and perturbed values). As shown in this article, see also the previous note~\cite{Cusin:2018ump}, since the physics on cosmological scale is quite well understood, the main source of uncertainties present in the predictions for the AGWB anisotropies is given by the description of galactic and sub-galactic physics.  This is extremely interesting because we are left with an observable that is very sensitive to the details of the astrophysical modeling as shown in Refs.~\cite{Cusin:2019jhg,Cusin:2019jpv}, and that can be used to constrain astrophysics.

The expressions that we provided in Refs.~\cite{Cusin:2017fwz, Cusin:2017mjm}, correspond to an astrophysical model where $\delta_{\cal L}=0$ in the Newtonian gauge \eqref{FL}. This is equivalent to assuming that the associated gauge invariant variable
$\delta_{\cal L} + (B-E')\partial \ln \bar{\cal L}_{\Gal}/\partial
\eta$ vanishes.  This is a very natural choice in the framework of the coarse
grained modeling used to describe anisotropies, since it allows one to compute both galaxy density and luminosity at the same time.  
Alternatively, in Ref.~\cite{Bertacca:2019fnt} the
modeling corresponds to the choice that the gauge invariant luminosity perturbation on
constant redshift hypersurfaces $\delta_{\cal L}^z  $vanishes
identically. The result of different choices for the $\delta_{\mathcal{L}}$ is expected to give little differences on observable scales since their difference (\ref{deltaZ}) is proportional to the time perturbation $\delta\eta=\eta_z-\eta$ and to the derivative of the galaxy luminosity with respect to time, and in any realistic astrophysical model the galaxy luminosity is expected to have a smooth time evolution. There are of course other physically reasonable models to set $\delta_{\cal L}$, and one could for instance assume that it
  vanishes in the synchronous gauge. One would then use the very simple
  expression~\eqref{Master3} in which the effect of metric perturbations would
  merely be interpreted as an integrated effect (with no physical projection effects present). 

\vspace{1 em}
\textit{Acknowledgements ---}  We are extremely grateful to R. Durrer
and P. Ferreira for a careful reading of this manuscript and for
valuable comments and to I. Dvorkin for valuable discussions.
This project has received funding from the European Research Council (ERC) under the European Union's Horizon 2020 research and innovation program (grant agreement No 693024). 
\bibliography{myrefs}

\end{document}